\documentclass[useAMS,usenatbib]{mn2e}

\usepackage{amsmath}
\usepackage{graphicx}
\usepackage{amssymb}

\title[Non-Gaussianity of WMAP data]
{Non-Gaussianity analysis on local morphological measures of WMAP data}

\author[Wiaux et al.]
{Y. Wiaux$^{1}$, P. Vielva$^{2}$, R. B. Barreiro$^{2}$, E. Mart{\'\i}nez-Gonz\'alez$^{2}$, P. Vandergheynst$^{1}$\\
$^{1}$Signal Processing Institute, Ecole Polytechnique F\'ed\'erale de Lausanne (EPFL), CH-1015 Lausanne, Switzerland\\
\hspace{0.1cm}E-mails : yves.wiaux@epfl.ch, pierre.vandergheynst@epfl.ch\\
$^{2}$Instituto de F{\'\i}sica de Cantabria (CSIC - UC), 39005 Santander, Spain\\
\hspace{0.1cm}E-mails : vielva@ifca.unican.es, barreiro@ifca.unican.es, martinez@ifca.unican.es\\}

\begin{document}

\date{\today}

\pagerange{\pageref{firstpage}--\pageref{lastpage}} \pubyear{2007}

\maketitle

\label{firstpage}

\begin{abstract}
The decomposition of a signal on the sphere with the steerable wavelet
constructed from the second Gaussian derivative gives access to the
orientation, signed-intensity, and elongation of the signal's local
features. In the present work, the non-Gaussianity of the WMAP temperature
data of the cosmic microwave background (CMB) is analyzed in terms
of the first four moments of the statistically isotropic random fields
associated with these local morphological measures, at wavelet scales
corresponding to angular sizes between $27.5'$ and $30^{\circ}$
on the celestial sphere. While no detection is made neither in the
orientation analysis nor in the elongation analysis, a strong detection
is made in the excess kurtosis of the signed-intensity of the WMAP
data. The non-Gaussianity is observed with a significance level below
$0.5\%$ at a wavelet scale corresponding to an angular size around
$10^{\circ}$, and confirmed at neighbour scales. This supports a
previous detection of an excess of kurtosis in the wavelet coefficient
of the WMAP data with the axisymmetric Mexican hat wavelet \citep{vielva04}.
Instrumental noise and foreground emissions are not likely to be at
the origin of the excess of kurtosis. Large-scale modulations of the
CMB related to some unknown systematics are rejected as possible origins
of the detection. The observed non-Gaussianity may therefore probably
be imputed to the CMB itself, thereby questioning the basic inflationary
scenario upon which the present concordance cosmological model relies.
Taking the CMB temperature angular power spectrum of the concordance
cosmological model at face value, further analysis also suggests that
this non-Gaussianity is not confined to the directions on the celestial
sphere with an anomalous signed-intensity.
\end{abstract}


\section{Introduction}

\label{sec:introduction}Among other cosmological observations, the
recent data of the cosmic microwave background, in particular those
released by the Wilkinson Microwave Anisotropy Probe (WMAP) satellite
experiment, have played a central role in defining a concordance model
highlighting a flat $\Lambda\textnormal{CDM}$ Universe with a primordial
phase of inflation \citep{bennett03,spergel03,hinshaw07,spergel07}.
In that framework, the flat Universe is filled in with cold dark matter
(CDM) and dark energy in the form of a cosmological constant ($\Lambda$),
in addition to the standard baryonic and electromagnetic components.
The cosmological principle is assumed, which postulates global homogeneity
and isotropy. In the basic inflationary scenario considered, the large
scale structure and the cosmic microwave background fluctuations are
also assumed to arise from Gaussian quantum energy density perturbations
around the homogeneous and isotropic background during the inflationary
phase. The cosmological parameters of the concordance cosmological
model are determined with a precision of the order of several percent,
but the hypotheses on which the model relies need to be thoroughly
challenged.

In that context, the CMB constitutes a realization of a random field
on the sphere. The cosmological principle and the basic inflationary
scenario respectively imply the statistical isotropy and the Gaussianity
of this random field. These basic hypotheses of the concordance cosmological
model may consequently be questioned through the CMB analysis.

The statistical isotropy of the CMB temperature field has already
been largely questioned in the analysis of the WMAP data. Firstly,
a North-South asymmetry in ecliptic coordinates has been detected
\citep{eriksen04a,eriksen04b,eriksen05,hansen04a,hansen04b,donoghue05,land05a,bernui06,bernui07,spergel07,eriksen07,monteserin07}.
Secondly, an anomalous alignment of the lowest multipoles of the data
was observed \citep{deOliveira04,schwarz04,copi04,katz04,bielewicz05,land05b,land07,freeman06,abramo06}.
Finally, wavelet analyses have also reported statistical isotropy
anomalies related to the signed-intensity of local CMB features, as
well as anomalies related to the alignment of local CMB features toward
specific directions on the celestial sphere \citep{wiaux06a,vielva06,vielva07}.

The Gaussianity of the CMB temperature field has also been largely
questioned in the analysis of the WMAP data. Firstly, departures from
Gaussianity were detected using statistics of extrema \citep{larson04,larson05,tojeiro06},
bispectra \citep{land05a}, phase correlations \citep{chiang03,chiang06,coles04,naselsky05},
Minkowski functionals \citep{park04,eriksen04b}, and local curvature
\citep{hansen04a}. Secondly, wavelet analyses have also reported
non-Gaussian deviations. An excess of kurtosis in the wavelet coefficient
of the WMAP temperature data with the axisymmetric Mexican hat wavelet
on the sphere was found at wavelet scales corresponding to angular
sizes on the celestial sphere around $10^{\circ}$, and localized
in the southern galactic hemisphere \citep{vielva04}. A cold spot
(\emph{i.e.} with negative wavelet coefficients) was identified at
$(\theta,\varphi)=(147^{\circ},209^{\circ})$, with $\theta\in[0,\pi]$
and $\varphi\in[0,2\pi)$ respectively standing for the co-latitude
and longitude in galactic spherical coordinates, and considered to
be a good candidate to explain the observed deviation. The confirmation
that the cold spot is anomalous was provided, still with the axisymmetric
Mexican hat wavelet, in terms of its area \citep{cruz05}. The detection
was further confirmed with various wavelets and various statistics
\citep{mukherjee04,cayon05,mcewen05,cruz06,cruz07b,mcewen06}. Notice
that the cold spot identified also certainly represents a departure
from statistical isotropy, in terms of a North-South asymmetry in
galactic coordinates.

By essence, wavelet analyses present the particular advantage of probing,
not only the scale but also the localization of the features constituting
the CMB on the celestial sphere \citep{wiaux05}. Steerable wavelets
also provide morphological measures of the local features, such as
orientation, signed-intensity, or elongation \citep{mcewen07}, at
a low computational cost \citep{wiaux06b}. They were used to probe
the statistical isotropy of the WMAP CMB temperature data in the previously
quoted signed-intensity and alignment analyses. They were also used
to probe the Integrated Sachs-Wolfe effect through the correlation
of WMAP data and large scale structure data \citep{mcewen07}. In
the present work, a further insight into the CMB temperature non-Gaussianity
is provided through a steerable wavelet analysis of the WMAP data.

In Section \ref{sec:methodology}, we present the methodology adopted.
In Section \ref{sec:analysis}, we present the results of the WMAP
data analysis. In Section \ref{sec:systematic}, we study systematic
effects as a possible origin of the detections. In Section \ref{sec:non-confinement},
we discuss the origin of our detection and its detailed interpretation.
We finally conclude in Section \ref{sec:conclusion}.

\section{Methodology}

\label{sec:methodology}In this section, we firstly recall the formalism
for the analysis of signals on the sphere with steerable wavelets,
as well as the local morphological measures of orientation, signed-intensity,
and elongation, defined from the steerable wavelet constructed from
the second Gaussian derivative. Secondly, we explicitly describe the
statistics for the non-Gaussianity analysis on the random fields associated
with the local morphological measures of the CMB temperature field.
These statistics are simply the first four moments of the random fields
considered.

\subsection{Steerable wavelets and morphological measures}

\label{sub:steerability-morphology}We consider the three-dimensional
Cartesian coordinate system $(o,o\hat{x},o\hat{y},o\hat{z})$ centered
on the unit sphere, and where the direction $o\hat{z}$ identifies
the North pole. Any point $\omega$ on the sphere is identified by
its corresponding spherical coordinates $(\theta,\varphi)$, where
$\theta\in[0,\pi]$ stands for the co-latitude, and $\varphi\in[0,2\pi)$
for the longitude.

Firstly, we briefly summarize the formalism of steerable wavelets
on the sphere $S^{2}$ \citep{wiaux05}. Any filter invariant under
rotation around itself is said to be axisymmetric. By definition,
any non-axisymmetric, or directional, filter $\Psi$ is steerable
if a rotation by $\chi\in[0,2\pi)$ around itself may be expressed
in terms of a finite linear combination of $M$ non-rotated basis
filters $\Psi_{m}$:\begin{equation}
\Psi_{\chi}\left(\omega\right)=\sum_{m=1}^{M}k_{m}\left(\chi\right)\Psi_{m}\left(\omega\right),\label{eq:steerability}\end{equation}
where the weights $k_{m}(\chi)$, with $1\leq m\leq M$, are called
interpolation functions. The analysis of a signal $F$ with a given
wavelet $\Psi$ simply defines a set of wavelet coefficients $W_{\Psi}^{F}(\omega_{0},\chi,a)$,
which result from the directional correlation between $F$ and the
wavelet dilated at any scale $a$, $\Psi_{a}$. In other words these
wavelet coefficients are defined by the scalar product between the
signal and the wavelet dilated at scale $a$, rotated around itself
by $\chi$, and translated at any point $\omega_{0}$ on the sphere,
also denoted $\Psi_{\omega_{0},\chi,a}$: \begin{equation}
W_{\Psi}^{F}(\omega_{0},\chi,a)=\langle\Psi_{\omega_{0},\chi,a}|F\rangle=\int_{S^{2}}d\Omega\Psi_{\omega_{0},\chi,a}^{*}(\omega)F(\omega).\label{eq:direccorr}\end{equation}
 The $^{*}$ denotes complex conjugation. The wavelet coefficients
of a signal therefore characterize the signal at each scale $a$,
orientation $\chi$, and position $\omega_{0}$. 

In the present work, we consider the second Gaussian derivative wavelet
(2GD), $\Psi^{\partial_{\hat{x}}^{2}(gau)}$, which is obtained by
a stereographic projection of the second derivative in direction $\hat{x}$
of a Gaussian in the tangent plane at the North pole. The filter obtained
is a steerable wavelet on the sphere which may be rotated in terms
of three basis filters ($M=3$): the second derivative in the direction
$\hat{x}$ itself, the second derivative in the direction $\hat{y}$,
and the cross-derivative. Notice that the value of the scale $a$
identifies with the dispersion of the Gaussian in units of $2\tan(\theta/2)$.
The angular size of the 2GD is defined as twice the half-width of
the wavelet, where the half-width is defined by $\theta_{hw}=2\arctan(a/2)$,
which is closely approximated by $a$ at small scales. 

Secondly, we recall that the 2GD gives access to three local morphological
measures of orientation, signed-intensity, and elongation \citep{mcewen07}.
By linearity, the relation of steerability (\ref{eq:steerability})
is automatically transferred on the wavelet coefficients of $F$.
Consequently, at each scale $a$ and at each position $\omega_{0}$,
the orientation $\chi_{0}(\omega_{0},a)$ that maximizes the absolute
value of the wavelet coefficient, can easily be computed, with an
infinite theoretical precision. It corresponds to the local orientation
at which the wavelet best matches the local feature of the signal.
As the 2GD is invariant under rotation around itself by $\pi$, orientations
may arbitrarily be constrained in a range of length $\pi$, and as
the 2GD oscillates in the tangent direction $\hat{x}$, it actually
detects features aligned along the tangent direction $\hat{y}$. The
local orientation of the feature itself, $D^{F}(\omega_{0},a)$, is
therefore defined in terms of $\chi_{0}=\chi_{0}(\omega_{0},a)$ as:
\begin{equation}
\frac{\pi}{2}\leq\quad D^{F}\left(\omega_{0},a\right)\equiv\chi_{0}+\frac{\pi}{2}\quad<\frac{3\pi}{2}.\label{eq:orientation}\end{equation}
The wavelet coefficient itself at scale $a$, position $\omega_{0}$,
and in direction $\chi_{0}$, defines to so-called signed-intensity
of the local feature:\begin{equation}
I^{F}\left(\omega_{0},a\right)\equiv W_{\Psi^{\partial_{\hat{x}}^{2}}}^{F}\left(\omega_{0},\chi_{0},a\right).\label{eq:signed-intensity}\end{equation}
The elongation of local features is explicitly defined by \begin{equation}
0\leq\quad E^{F}\left(\omega_{0},a\right)\equiv1-\Biggl\vert\frac{W_{\Psi^{\partial_{\hat{x}}^{2}}}^{F}\left(\omega_{0},\chi_{0}+\frac{\pi}{2},a\right)}{W_{\Psi^{\partial_{\hat{x}}^{2}}}^{F}\left(\omega_{0},\chi_{0},a\right)}\Biggl\vert\quad\leq1.\label{eq:elongation}\end{equation}
Numerical tests performed on elliptical Gaussian-profile features
show that this elongation measure increases monotonously in the range
$[0,1]$ with the intrinsic eccentricity $e\in[0,1]$ of the features.
While it is possible to define alternative elongation measures, these
numerical tests also indicate that the chosen definition is not an
arbitrary measure of the non-axisymmetry of local features, but represents
a rough estimate of the eccentricity of a Gaussian-profile local feature. 

In summary, the analysis of signals with steerable wavelets is interesting
in several respects. Firstly, the wavelet decomposition enables one
to identify the scales $a$ of the physical processes which define
the local feature of the signal at each point $\omega_{0}$. Secondly,
the steerability theoretically gives access to local morphological
measures. For the 2GD, the orientation, signed-intensity and elongation
of local features are defined. Finally, from the computational point
of view, the calculation of a directional correlation at each analysis
scale is an extremely demanding task. The relation of steerability
is essential to reduce the complexity of calculation of the wavelet
coefficients when local orientations are considered \citep{wiaux06b}.

\subsection{Statistics for non-Gaussianity}

\label{sub:Statistics-for-non-Gaussianity}In the context of the concordance
cosmological model, the CMB temperature represents a realization of
a statistically isotropic and Gaussian random field on the sphere.
The WMAP data are also contaminated by noise and foreground emissions.
The statistical analysis is performed by comparison of the data with
simulations. The noise present in the data is simulated and the regions
of the sky in which the data are too much contaminated by foreground
emissions are masked, and excluded from the analysis. Typically, a
non-Gaussianity analysis is performed through the evaluation of global
estimators computed as simple averages on the whole part of the celestial
sphere where the data are considered to be valid, explicitly assuming
the statistical isotropy in the corresponding part of the sky. Any
anomaly between the data and the simulations is consequently interpreted
as a departure of the data from Gaussianity.

We consider the statistically isotropic real-valued random fields
on the sphere associated with the local morphological measures of
orientation, signed-intensity, and elongation of the CMB temperature
field $T$ at each wavelet scale $a$: $X(\omega_{0},a)$, with $X=\{D^{T},I^{T},E^{T}\}$.
The statistics estimated for the subsequent non-Gaussianity analysis
are simply moments of the first four orders. The first two are the
mean $\mu^{X}(a)=\langle X(\omega_{0},a)\rangle$, and variance $\sigma^{X}(a)=\langle[X(\omega_{0},a)-\mu^{X}(a)]^{2}\rangle^{1/2}$.
The third-order moment is the skewness $S^{X}(a)=\langle[X(\omega_{0},a)-\mu^{X}(a)]^{3}\rangle/[\sigma^{X}(a)]^{3}$.
The skewness measures the asymmetry of the probability density function,
and hence a deviation relative to a Gaussian distribution. Positive
and negative skewnesses are respectively associated with larger right
and left distribution tails. The fourth-order moment considered is
the excess kurtosis $K^{X}(a)=\langle[X(\omega_{0},a)-\mu^{X}(a)]^{4}\rangle/[\sigma^{X}(a)]^{4}-3$.
The kurtosis measures the peakedness of the probability density function
relative to a Gaussian distribution. Positive and negative excess
kurtoses are respectively associated with distributions more and less
peaked than a Gaussian distribution. These four moments are independent
of the point $\omega_{0}$ because of the statistical isotropy. The
corresponding estimators computed by averages over the sphere are\begin{eqnarray}
\widehat{\mu}^{X}\left(a\right) & = & \frac{1}{N_{a}}\sum_{i=1}^{N_{a}}X\left(\omega_{0}^{(i)},a\right)\nonumber \\
\widehat{\sigma}^{X}\left(a\right) & = & \left\{ \frac{1}{N_{a}}\sum_{i=1}^{N_{a}}\left[X\left(\omega_{0}^{(i)},a\right)-\widehat{\mu}^{X}\left(a\right)\right]^{2}\right\} ^{1/2}\nonumber \\
\widehat{S}^{X}\left(a\right) & = & \frac{1}{N_{a}}\sum_{i=1}^{N_{a}}\left[\frac{X\left(\omega_{0}^{(i)},a\right)-\widehat{\mu}^{X}\left(a\right)}{\widehat{\sigma}^{X}\left(a\right)}\right]^{3}\nonumber \\
\widehat{K}^{X}\left(a\right) & = & \frac{1}{N_{a}}\sum_{i=1}^{N_{a}}\left[\frac{X\left(\omega_{0}^{(i)},a\right)-\widehat{\mu}^{X}\left(a\right)}{\widehat{\sigma}^{X}\left(a\right)}\right]^{4}-3.\label{eq:statistics}\end{eqnarray}
At each wavelet scale $a$, $N_{a}$ stands for the total number of
valid pixels outside a given exclusion mask $M_{a}$ which, by definition,
identifies the pixels to be excluded from the analysis (see Subsection
\ref{sub:data-simulations}). The values $\omega_{0}^{(i)}$ identify
the center of these valid pixels.

Let us emphasize here that, even under the assumption of Gaussianity
of the CMB temperature field, postulated for the simulations, none
of the random fields associated with these local morphological measures
at each scale is intrinsically Gaussian. This is simply due to the
non-linearity of the definitions (\ref{eq:orientation}) for the orientation,
(\ref{eq:signed-intensity}) for the signed-intensity, and (\ref{eq:elongation})
for the elongation. In particular, by statistical isotropy, the measure
of orientation $D^{F}(\omega_{0},a)$ should be uniformly distributed
at each point $\omega_{0}$ and at each wavelet scale $a$.

For each local morphological measure, statistics, and wavelet scale,
the value obtained for the data can be compared to the corresponding
values for the simulations. The percentiles corresponding to specific
cumulative probabilities $p$ in the simulations considered are calculated
for a first comparison with the value of the data. The percentile
associated with $p=50\%$ defines the median value. Cumulative probabilities
$p=\{15.865\%,84.135\%\}$ are considered, which formally correspond
to the percentiles at one standard deviation ($1\sigma$) from the
mean in a Gaussian distribution. They define a first, innermost, region
for the distribution of percentiles around the median value. The exact
values considered reflect the maximum precision allowed by our sample
of ten thousand simulations. Cumulative probabilities $p=\{2.275\%,97.725\%\}$
are also considered, which formally correspond to the percentiles
at two standard deviations ($2\sigma$) from the mean in a Gaussian
distribution. They define a second, middle, region for the distribution
of percentiles around the median value. Again, the exact values considered
reflect the maximum precision allowed by our sample of ten thousand
simulations. Cumulative probabilities $p=\{0.5\%,99.5\%\}$ are finally
considered, defining a third, outermost, region for the distribution
of percentiles around the median value. If the value of the data for
a statistics at a given scale is higher (lower) than the median value
obtained from the simulations, the significance level of a detection
is simply defined as the fraction of simulations with a higher (lower)
value than the data. The lower the significance level, the stronger
the detection. Typically, in the following, a significance level below
$0.5\%$, corresponding to values outside the outermost region for
the distribution of percentiles around the median value, will be associated
with a strong detection.

\section{WMAP analysis}

\label{sec:analysis}In this section, we firstly describe the pre-processing
procedure applied to the three-year WMAP CMB temperature data, as
well as to the corresponding simulations produced from the concordance
cosmological model. Secondly, we expose the results of the application
of the non-Gaussianity analysis defined in the previous section on
the three-year WMAP co-added CMB data, notably highlighting a strong
detection in the excess kurtosis of the signed-intensity.

\subsection{Data and simulations}

\label{sub:data-simulations}Firstly, the following pre-processing
procedure is applied to the three-year WMAP CMB temperature data before
the non-Gaussianity analysis. The original maps of the eight WMAP
radiometers at the Q, V, and W frequencies (Q1 and Q2 at $41$ GHz,
V1 and V2 at $61$ GHz, and W1, W2, W3, and W4 at $94$ GHz) are corrected
for foreground emissions contamination by a template fitting technique
\citep{spergel07}. The resulting foreground cleaned maps are available
from the NASA LAMBDA archive%
\footnote{http://lambda.gsfc.nasa.gov/%
}. These maps are masked with the Kp0 mask \citep{spergel07} that
cuts the regions of brightest galactic emission around the galactic
plane ($\approx20\%$ of the sky), as well as the brightest galactic
and extragalactic point sources ($\approx5\%$ of the sky). Zero values
are assigned to the corresponding pixels. The instrumental beam associated
with the WMAP radiometers is described by an isotropic window function,
and the instrumental noise is to first order Gaussian, statistically
anisotropic, and uncorrelated. A map with better signal-to-noise ratio
can be obtained by an optimal combination of the eight foreground
cleaned and masked maps. At each pixel, this combination is obtained
by weighting each map by the corresponding inverse noise variance.
In order to minimize any error coming from the cosmological dipole
subtraction, the dipole outside the mask is removed \citep{komatsu03}.
This overall pre-processing procedure defines the so-called three-year
WMAP co-added CMB map \citep{hinshaw07}, which is used in the subsequent
analysis. 

Secondly, ten thousand simulations of the three-year WMAP co-added
CMB map are considered to compare the results of the analysis of the
data to what is expected from the concordance model. Each simulation
is produced as follows. Spherical harmonics coefficients of a statistically
isotropic and Gaussian CMB realization are obtained from the angular
power spectrum determined by the cosmological parameters of the three-year
WMAP best-fit model \citep{spergel07} with CAMB (Code for Anisotropies
in the Microwave Background%
\footnote{http://camb.info/%
}). The observation at each of the eight WMAP radiometers of the Q,
V, and W frequencies is simulated by convolving that realization in
harmonic space with the corresponding isotropic window function. Each
map is then transformed to pixel space at the appropriate resolution,
and a Gaussian, statistically anisotropic, and uncorrelated noise
realization is added with the proper variance per pixel. This provides
simulations of the CMB, as seen by the eight radiometers at the different
WMAP frequencies considered. The same prescriptions as those described
above for the data are then applied to produce a three-year WMAP co-added
CMB map.

Notice that the WMAP co-added CMB maps for the data and simulations
are initially produced in HEALPix pixelization%
\footnote{http://healpix.jpl.nasa.gov/%
} \citep{gorski05} at the resolution $N_{side}=512$, corresponding
to maps with more than three million equal-area pixels with a spatial
resolution of $6.87'$. For the sake of our analysis, which is applied
at $17$ scales of the 2GD wavelet, corresponding to angular sizes
between $27.5'$ and $30^{\circ}$, the maps are downgraded to the
resolution $N_{side}=256$. This provides maps with a bit less than
one million equal-area pixels with a spatial resolution of $13.7'$.
Also notice that, in pixels close to masked regions, the result of
the directional correlation of a signal with a steerable wavelet is
inevitably affected by the zero values of the Kp0 mask. An exclusion
mask $M_{a}$ is therefore defined at each wavelet scale $a$, identically
on the data and simulations, in order to exclude the affected pixels
from the analysis \citep{vielva04}, leaving $N_{a}$ valid pixels
from which statistics may be estimated.

\subsection{Non-Gaussianity analysis}

\label{sub:non-Gaussianity-analysis}%
\begin{figure}
\begin{center}\includegraphics[width=8cm,height=6cm]{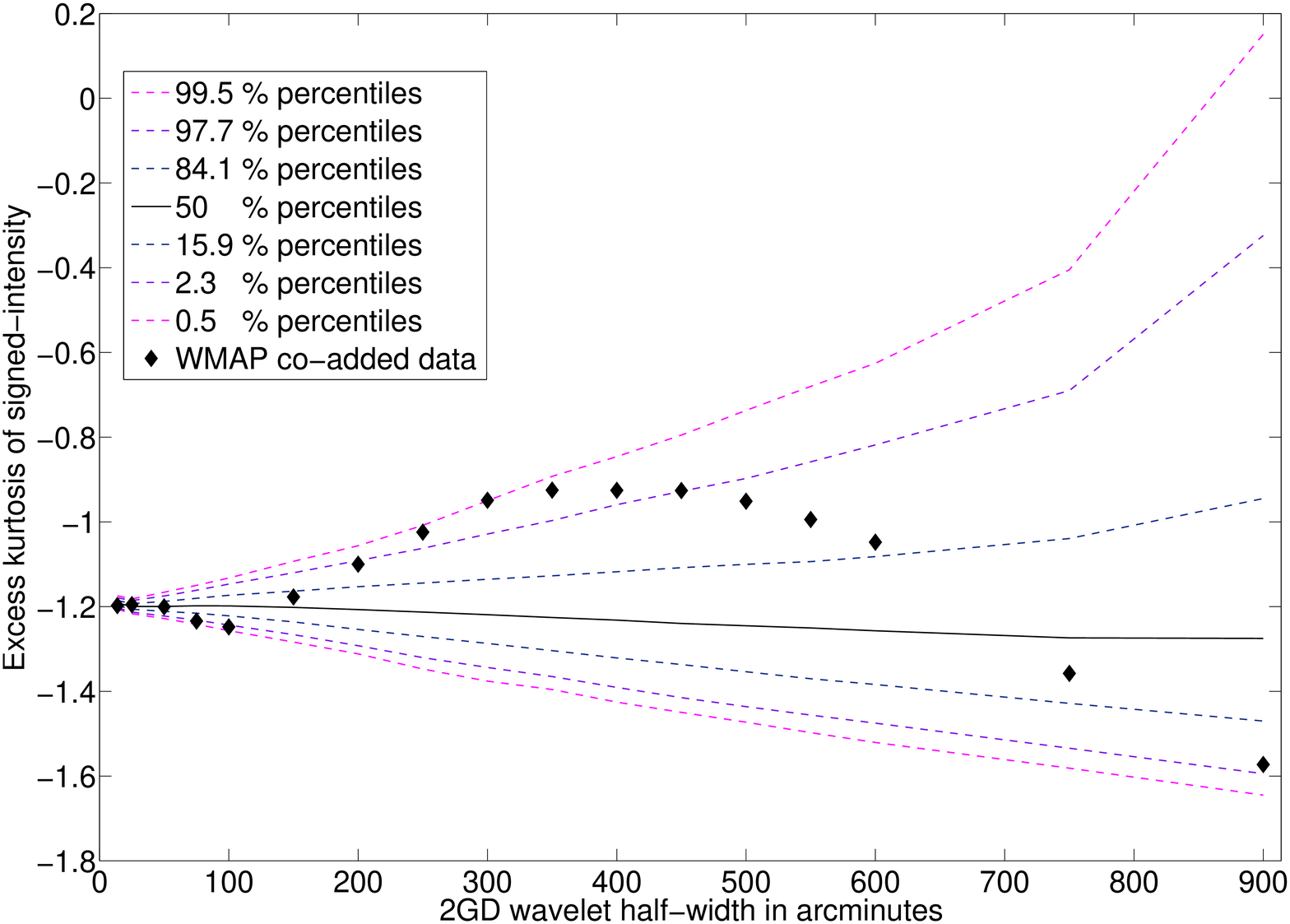}\end{center}

\caption{\label{fig:kurtosis-scales1-17}Excess kurtosis of the signed-intensity
of the three-year WMAP co-added CMB data as a function of the 2GD
wavelet half-width in a range corresponding to angular sizes between
$27.5'$ and $30^{\circ}$ on the celestial sphere. Data (black rhombi)
are compared with percentiles established from ten thousand statistically
isotropic and Gaussian simulations produced from the concordance cosmological
model. Significance levels lie roughly below $1.4\%$ at the four
wavelet scales $a_{8}$, $a_{9}$, $a_{10}$, and $a_{11}$, respectively
corresponding to angular sizes of $8.33^{\circ}$, $10^{\circ}$,
$11.7^{\circ}$, and $13.3^{\circ}$. The significance level reaches
a minimum value of $0.49\%$ at scale $a_{9}$. This identifies a
strong detection of non-Gaussianity, in terms of an excess of kurtosis
in the signed-intensity. }

\end{figure}
\begin{figure}
\begin{center}\includegraphics[width=8cm,height=6cm]{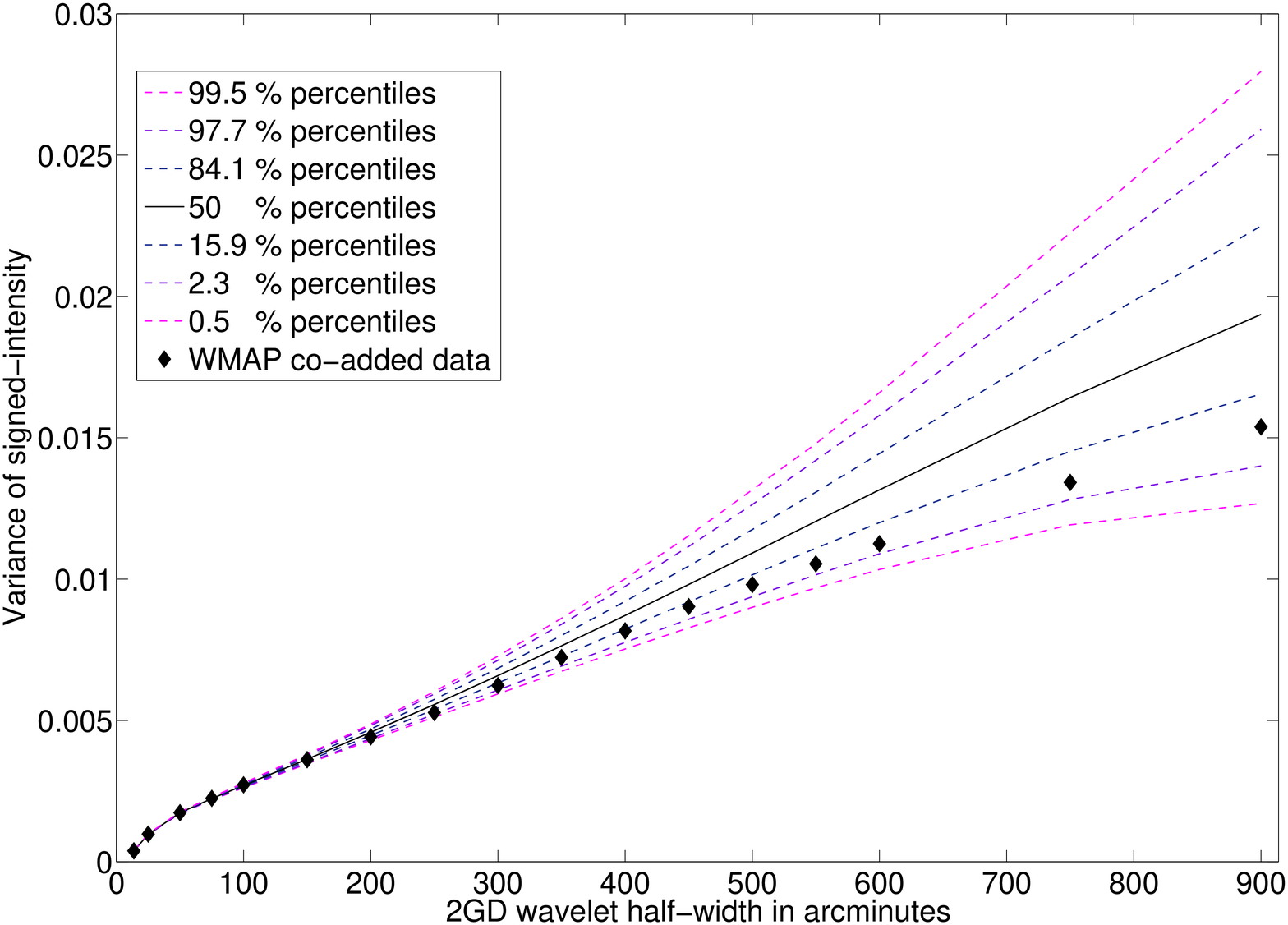}\end{center}

\caption{\label{fig:variance-scales1-17}Variance of the signed-intensity of
the three-year WMAP co-added CMB data as a function of the 2GD wavelet
half-width in a range corresponding to angular sizes between $27.5'$
and $30^{\circ}$ on the celestial sphere. Data (black rhombi) are
compared with percentiles established from ten thousand statistically
isotropic and Gaussian simulations produced from the concordance cosmological
model. Significance levels lie roughly above $5\%$ between the wavelet
scales $a_{2}$ and $a_{17}$, corresponding to angular sizes between
$50'$ and $30^{\circ}$. At the wavelet scale $a_{1}$ though, corresponding
to an angular size of $27.5'$, the significance level reaches $0\%$.
This identifies a strong, but isolated detection of non-Gaussianity,
in terms of a too high variance of the signed-intensity. }

\end{figure}
The results of the application of the non-Gaussianity analysis of
the local orientation, signed-intensity, and elongation to the three-year
WMAP co-added CMB map are as follows. Let us recall that $17$ scales
of the 2GD wavelet are probed, corresponding to angular sizes on the
celestial sphere between $27.5'$ and $30^{\circ}$. The complete
list of the wavelet half-widths $\theta_{hw}$ considered in arcminutes
reads: \{$13.7'$, $25'$, $50'$, $75'$, $100'$, $150'$, $200'$,
$250'$, $300'$, $350'$, $400'$, $450'$, $500'$, $550'$, $600'$,
$750'$, $900'$\}.

Firstly, the significance levels observed for statistics of the local
orientation and of the local elongation reach minimum values between
$5\%$ and $10\%$, and only at some isolated scales. In other words,
these values lie well inside the middle region defined above for the
distribution of percentiles around the median value. We therefore
conclude that no detection is obtained neither for the local orientation
nor for the local elongation.

Secondly, the significance levels observed for the excess kurtosis
of the signed-intensity are roughly below $1.4\%$ at the four wavelet
scales $a_{8}$, $a_{9}$, $a_{10}$, and $a_{11}$, respectively
corresponding to angular sizes of $8.33^{\circ}$, $10^{\circ}$,
$11.7^{\circ}$, and $13.3^{\circ}$ on the celestial sphere. The
excess kurtosis in the signed-intensity of the data at each of these
wavelet scales is higher than the median value defined by the simulations.
The significance level reaches a minimum value of $0.49\%$ at scale
$a_{9}$. These results identify a strong detection of non-Gaussianity
in the WMAP co-added CMB data, in terms of an excess of kurtosis in
the signed-intensity (see Figure \ref{fig:kurtosis-scales1-17}).
Also notice significance levels of roughly $1.6\%$ and $1.2\%$,
respectively at the wavelets scales $a_{4}$ and $a_{5}$, corresponding
to angular sizes of $2.5^{\circ}$ and $3.33^{\circ}$. The excess
kurtosis in the signed-intensity of the data at each of these wavelet
scales is well lower than the median value defined by the simulations,
even though it is not considered as anomalous.

Thirdly, the significance levels observed for the variance of the
signed-intensity are roughly above $5\%$ between the wavelet scales
$a_{2}$ and $a_{17}$, corresponding to angular sizes between $50'$
and $30^{\circ}$, the minimum value being reached at scale $a_{8}$.
Again, these values lie well inside the middle region defined above
for the distribution of percentiles around the median value. We therefore
conclude that no detection is obtained at those scales. At the wavelet
scale $a_{1}$ though, corresponding to an angular size of $27.5'$,
the value of the variance is well higher than the median value defined
by the simulations, and the significance level actually reaches $0\%$.
Formally, this represents a strong, but isolated detection of non-Gaussianity
in the WMAP co-added CMB data, in terms of a too high variance of
the signed-intensity (see Figure \ref{fig:variance-scales1-17}).
In Section \ref{sec:systematic}, we suggest that it might originate
in the presence of residual point sources in the data, and that it
should therefore be discarded. Notice that no detection appears neither
in the mean nor in the skewness of the signed-intensity.

In summary, the 2GD wavelet gives access to the measures of orientation,
signed-intensity, and elongation of local features of the WMAP temperature
data. But a strong detection of non-Gaussianity is only observed in
the excess kurtosis of the signed-intensity. This result actually
supports the previous detection, with the axisymmetric Mexican hat
wavelet, of an excess of kurtosis in the wavelet coefficient of the
WMAP temperature data \citep{vielva04,mukherjee04,cruz05}.

\section{Systematic effects}

\label{sec:systematic}In this section, we firstly suggest that the
high variance of the signed-intensity of the WMAP temperature data
at the smallest wavelet scales, and the corresponding isolated detection,
might originate in the presence of residual point sources in the data.
Secondly, we discard instrumental noise, residual foreground emissions,
and large-scale modulations of the CMB temperature field related to
some unknown systematics, as possible origins of the excess of kurtosis
in the signed-intensity.

\subsection{Residual point sources and lower resolution}

\label{sub:point-sources}The variance of the signed-intensity of
the WMAP co-added CMB data is lower than the median value defined
by the simulations between the wavelet scales $a_{6}$ and $a_{17}$,
corresponding to angular sizes between $5^{\circ}$ and $30^{\circ}$.
This variance is not considered as anomalous though, as significance
levels are always roughly above $5\%$. On the contrary, the variance
observed between the wavelet scales $a_{1}$ and $a_{5}$, corresponding
to angular sizes between $27.5'$ and $3.33^{\circ}$, is higher than
the median value. Again, the values of the variance between the wavelet
scales $a_{2}$ and $a_{5}$ are not considered as anomalous, as significance
levels are always roughly above $6\%$. As already emphasized, at
the wavelet scale $a_{1}$ , the significance level reaches $0\%$.
The high variance at the smallest wavelet scales, in opposition with
the behaviour at larger wavelet scales (see Figure \ref{fig:variance-scales1-17}),
might originate in systematic effects such as the presence of residual
point sources. Let us recall that the three-year WMAP best-fit angular
power spectrum is obtained after correction for a non-zero best-fit
amplitude of a residual point sources power spectrum \citep{hinshaw07}.
Consistently, residual point sources were recently identified in the
WMAP co-added CMB data \citep{lopezcaniego07}. These residual point
sources are however not accounted for in the WMAP co-added CMB data
analyzed here, while they are accounted for in the simulations based
on the three-year WMAP best-fit angular power spectrum. This could
indeed explain the high variance observed in the signed-intensity
of the data at the smallest wavelet scales. Obviously, the contribution
of these residual point sources is negligible at larger wavelet scales. 

In the absence of a detailed model of the contribution of the residual
point sources to the variance of the signed-intensity, the corresponding
isolated detection at the wavelet scale $a_{1}$ is discarded, and
the analysis is restricted to the wavelet scales between $a_{6}$
and $a_{17}$, corresponding to angular sizes between $5^{\circ}$
and $30^{\circ}$. This does not affect the interpretation of the
detection relative to the excess kurtosis in the signed-intensity,
which appears between the wavelet scales $a_{8}$ and $a_{11}$, corresponding
to angular sizes between $8.33^{\circ}$ and $13.3^{\circ}$. For
the subsequent analyses dedicated to search for the origin of the
detection in the excess kurtosis of the signed-intensity, the three-year
WMAP co-added CMB map and the corresponding ten thousand simulations
are downgraded to the resolution $N_{side}=32$. This provides maps
with a bit more than twelve thousand equal-area pixels with a spatial
resolution of $1.83^{\circ}$. For coherence, the initial non-Gaussianity
analysis is reproduced for the signed-intensity, from the data and
simulations preliminary downgraded to the resolution $N_{side}=32$.
All conclusions relative to the first four statistical moments remain
obviously unchanged. The strong detection in the excess kurtosis is
slightly enhanced. The corresponding significance levels are roughly
below $1\%$ at the four wavelet scales $a_{8}$, $a_{9}$, $a_{10}$,
and $a_{11}$, respectively corresponding to angular sizes of $8.33^{\circ}$,
$10^{\circ}$, $11.7^{\circ}$, and $13.3^{\circ}$ on the celestial
sphere. The significance level reaches a minimum value of $0.26\%$
at scale $a_{8}$.

\subsection{Other systematics}

\label{sub:other-systematics}Let us recall the previous detection
of an excess of kurtosis in the wavelet coefficient of the WMAP temperature
data with the axisymmetric Mexican hat wavelet \citep{vielva04,mukherjee04,cruz05},
as well as previous detections of anomalies obtained in the signed-intensity
of the WMAP temperature data with the 2GD wavelet \citep{mcewen07,vielva07}.
The corresponding analyses concluded that neither instrumental noise
nor residual foreground emissions are at the origin of the deviations
observed. These conclusions were drawn from independent analyses of
the data produced by the eight WMAP radiometers at the Q, V, and W
frequencies. These results suggest that the wavelets and statistics
used are rather insensitive to instrumental noise and residual foreground
emissions in the WMAP temperature data. In this context, even though
no similar analysis is performed here, we conclude that neither instrumental
noise nor residual foreground emissions are likely to be at the origin
of the excess of kurtosis observed in the signed-intensity of the
WMAP temperature data.

Further investigations are proposed here, considering some possible
form of unknown systematics. It was recently proposed that the WMAP
data are possibly affected by a large-scale modulation. This modulation
was primarily put forward as a possible explanation of the North-South
asymmetry and low multipoles alignment of the WMAP data \citep{helling06,gordon07}.
In that framework, the WMAP data are of the form $T(\omega)\times[1+f(\omega)]$,
where $T(\omega)$ stands for the CMB temperature on the sky, and
$f(\omega)$ is a modulation function containing only low multipoles
$l$. Dipolar ($l=1$) and dipolar-quadrupolar ($l=\{1,2\}$) modulation
functions providing the best-fit cosmological models to the three-year
WMAP data were proposed. Let us remark that the best-fit dipolar modulation
used \citep{eriksen07}, as well as the best-fit dipolar-quadrupolar
modulation used \citet[see arXiv:astro-ph/0603449v1]{spergel07},
were not primarily computed for the three-year WMAP co-added CMB map
itself, but they were shown not to be sensitive to the three-year
WMAP data set and sky cut. We therefore also considered them to be
adequate for correction of the three-year WMAP co-added CMB map.

We have checked the stability of the excess of kurtosis in the signed-intensity
of the three-year WMAP co-added CMB data relative to these modulations.
Firstly, considering the best-fit dipolar-quadrupolar modulation,
the strong detection in the excess kurtosis remains unchanged, if
it is not slightly increased. Just as for the analysis of the non-corrected
three-year WMAP co-added CMB data, the significance levels are roughly
below $1\%$ at the four wavelet scales $a_{8}$, $a_{9}$, $a_{10}$,
and $a_{11}$, respectively corresponding to angular sizes of $8.33^{\circ}$,
$10^{\circ}$, $11.7^{\circ}$, and $13.3^{\circ}$ on the celestial
sphere. The significance levels reach a minimum value of $0.22\%$
at scales $a_{8}$ and $a_{9}$. Secondly, considering the best-fit
dipolar modulation, the detection in the excess kurtosis is slightly
decreased. The significance levels are only roughly below $3\%$ at
the four wavelet scales $a_{8}$, $a_{9}$, $a_{10}$, and $a_{11}$,
with a minimum value of $0.60\%$ at scale $a_{8}$. As already emphasized
\citep{vielva07}, a more precise definition of the modulation in
terms of specific systematic effects would be required before strong
conclusions can be drawn from the application of the corresponding
corrections. But, even taking the results at face value, the proposed
dipolar and dipolar-quadrupolar corrections are to be rejected as
possible origins of the observed excess of kurtosis in the signed-intensity
of the WMAP temperature data.

In summary, instrumental noise and residual foreground emissions are
not likely to be at the origin of the excess of kurtosis. Large-scale
modulations of the CMB related to some unknown systematics are explicitly
rejected as possible origins of the detection. The non-Gaussianity
detected in the excess kurtosis of the signed-intensity of the WMAP
data is therefore probably related to the CMB temperature field itself.

\section{Confinement and discussion}

\label{sec:non-confinement}We here firstly recall the recent detection
of an anomalous distribution on the sky of anomalous signed-intensities
in the three-year WMAP co-added CMB data. We secondly test, and tend
to reject, the possible confinement of the observed excess of kurtosis
to the directions with an anomalous signed-intensity. We finally discuss
the detailed interpretation of our detections.

\subsection{Local signed-intensity anomalies}

\label{sub:local-signed-intensity-anomalies}In a very recent analysis
of the three-year WMAP co-added CMB data with the 2GD wavelet, the
distribution on the celestial sphere of directions with a signed-intensity
anomalous at $99.865\%$ (formally corresponding to the percentiles
at three standard deviations ($3\sigma$) from the mean in a Gaussian
distribution) was observed to be anomalous \citep{vielva07}. At the
wavelet scale $a_{8}$, corresponding to an angular size of $8.33^{\circ}$,
the global significance level of that detection, defined as the fraction
of the ten thousand simulations with a number of anomalous directions
higher than in the data, is $1.39\%$. The anomalous directions are
essentially distributed in three clusters in the southern galactic
hemisphere, identifying three mean preferred directions in the sky
\citet[Figure 5]{vielva07}. A first cold spot (\emph{i.e.} with negative
signed-intensities) identifies with the anomalous cold spot originally
detected at $(\theta,\varphi)=(147^{\circ},209^{\circ})$ in galactic
spherical coordinates with the axisymmetric Mexican hat wavelet \citep{vielva04,cruz05}.
A second cold spot lies very close to the southern end of the CMB
dipole axis. The third spot is a hot spot (\emph{i.e.} with positive
signed-intensities) close to the southern end of the ecliptic poles
axis. The detection is confirmed at the neighbour wavelet scales $a_{9}$,
$a_{10}$, and $a_{11}$, respectively corresponding to angular sizes
of $10^{\circ}$, $11.7^{\circ}$, and $13.3^{\circ}$. Instrumental
noise, residual foreground emissions, as well as large-scale modulations
of the CMB related to some unknown systematics, are rejected as possible
origins of the detection. The localized anomalous distribution of
anomalous signed-intensities identified may therefore probably be
imputed to the CMB temperature field itself.

\subsection{Confinement analysis}

\label{sub:confinement-analysis}%
\begin{figure}
\begin{center}\includegraphics[width=8cm,height=6cm]{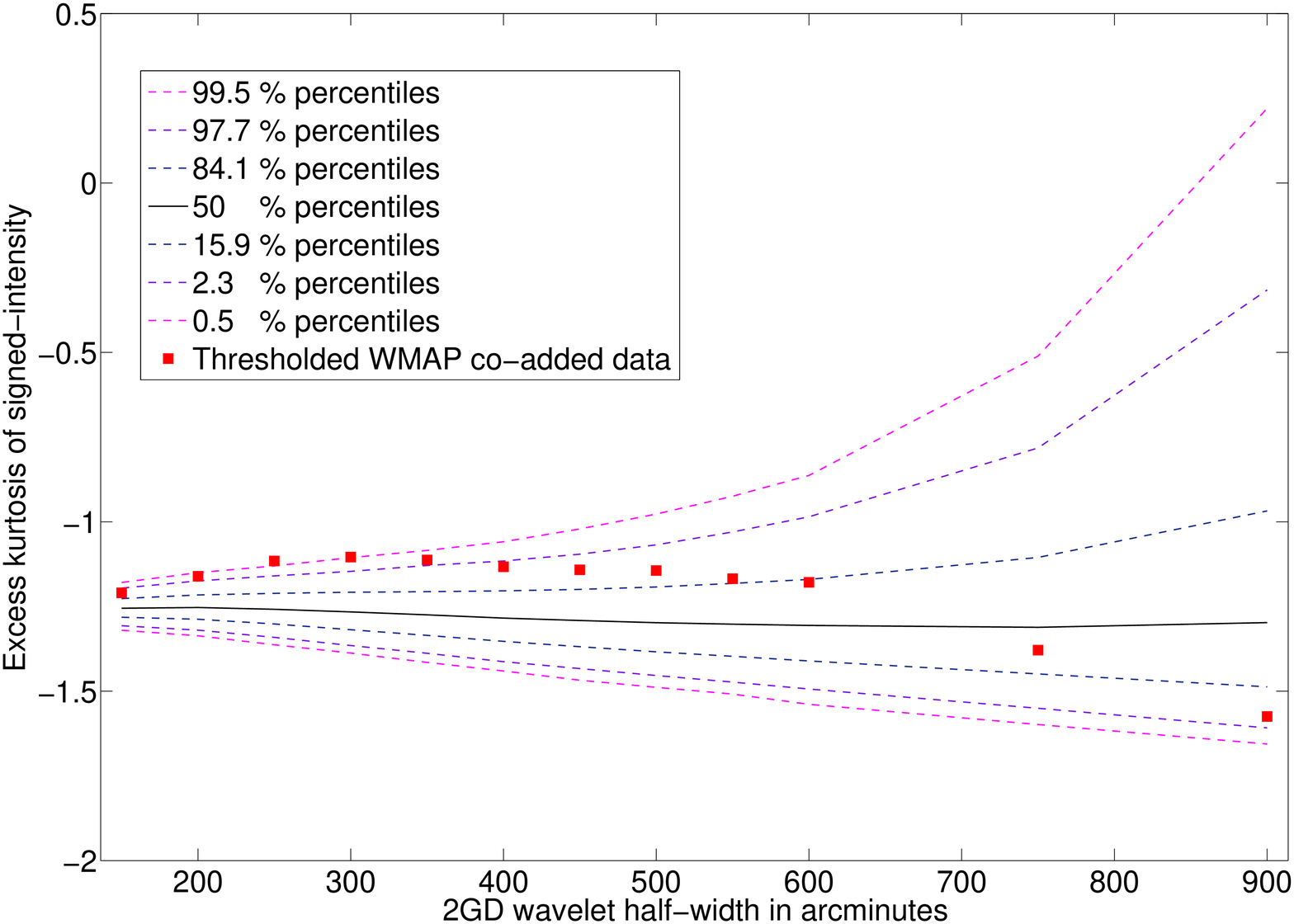}\end{center}

\caption{\label{fig:adaptivethres-kurtosis-scales6-17}Excess kurtosis of the
signed-intensity of the three-year WMAP co-added CMB data as a function
of the 2GD wavelet half-width in a range corresponding to angular
sizes between $5^{\circ}$ and $30^{\circ}$ on the celestial sphere.
Data (red squares) are compared with percentiles established from
ten thousand statistically isotropic and Gaussian simulations produced
from the concordance cosmological model. The directions with an anomalous
signed-intensity are excluded from the statistical analysis, in the
data as well as in each of the ten thousand simulations independently.
Significance levels are roughly below $1\%$ at the four wavelet scales
$a_{7}$, $a_{8}$, $a_{9}$, and $a_{10}$, respectively corresponding
to angular sizes of $6.67^{\circ}$, $8.33^{\circ}$, $10^{\circ}$,
and $11.7^{\circ}$. The significance level reaches a minimum value
of $0.28\%$ at scale $a_{8}$. This still identifies a strong detection
of non-Gaussianity, in terms of an excess of kurtosis in the signed-intensity. }

\end{figure}
Postulating the Gaussianity of the CMB, one may interpret the anomalous
distribution of directions with an anomalous signed-intensity as a
clear departure from statistical isotropy \citep{vielva07}. But more
generally, the anomaly observed highlights a deviation of the CMB
temperature field from the whole assumption of statistical isotropy
and Gaussianity. In the context of the present non-Gaussianity detection,
the anomalous distribution of anomalous signed-intensities previously
identified represents a serious candidate to explain the excess of
kurtosis observed in the signed-intensity of the three-year WMAP co-added
CMB data. This idea is supported by the fact that the two detections
are observed at the same wavelet scales. The hypothesis to check is
to know if the whole non-Gaussianity observed is confined to the directions
with an anomalous signed-intensity, in the idea that the excess in
the number of anomalous directions would bear the departure from Gaussianity.

The confinement analysis simply consists in reproducing the previous
analysis on the first four statistical moments of the signed-intensity
of the three-year WMAP co-added CMB data, from which the directions
with an anomalous signed-intensity are excluded. The only way to account
for a possible bias introduced by the exclusion of the extremal values
of the data above a given threshold is to apply the same exclusion
process to each simulation independently. In particular, the coherence
of the procedure can only be achieved if all values above the threshold,
and not only part of it, are identified and excluded from the statistical
analysis, in the data and in the simulations%
\footnote{We therefore have to assume that, in the data, all extremal values
of the background Gaussian distribution above the threshold are excluded.
The only formal reason for which extremal values outside the effectively
observed directions with a signed-intensity above the threshold might
have been missed would be that some non-Gaussianity coincidentally
compensates for the extremal value in the corresponding direction
on the celestial sphere. In that trivial case, one readily knows that
the non-Gaussianity observed in the WMAP data is not confined to the
localized distribution of anomalous signed-intensities.%
}. Notice that, while the anomalous signed-intensities were originally
identified at a threshold of $99.865\%$, we consider here a threshold
at $99.5\%$. This lowering is performed in order to avoid a possible
negative conclusion, relative to the confinement, simply due to the
fact that directions in the immediate vicinity of the directions thresholded
are not excluded from the analysis. For completeness, even though
the detections are only observed at the four wavelet scales between
$a_{8}$ and $a_{11}$, corresponding to angular sizes between $8.33^{\circ}$
and $13.3^{\circ}$, the confinement analysis is performed at all
wavelet scales between $a_{6}$ and $a_{17}$, corresponding to angular
sizes between $5^{\circ}$ and $30^{\circ}$.

The results of the analysis performed are as follows. The strong detection
of an excess of kurtosis in the signed-intensity of the WMAP temperature
data is preserved. The significance levels are roughly below $1\%$
at the four wavelet scales $a_{7}$, $a_{8}$, $a_{9}$, and $a_{10}$,
respectively corresponding to angular sizes of $6.67^{\circ}$, $8.33^{\circ}$,
$10^{\circ}$, and $11.7^{\circ}$ on the celestial sphere. The significance
level reaches a minimum value of $0.28\%$ at scale $a_{8}$ (see
Figure \ref{fig:adaptivethres-kurtosis-scales6-17}). The variance
of the signed-intensity at each wavelet scale was well lower than
the median value defined by the simulations before the thresholding.
It was not considered as anomalous as the significance level reached
a minimum value of roughly $5\%$ at scale $a_{8}$. The variance
of the signed-intensity at each wavelet scale is still well lower
than the median value defined by the simulations after the thresholding.
It is still not considered as anomalous, even though the significance
level reaches a minimum value of roughly $1\%$ at scale $a_{8}$.
No detection appears neither in the mean nor in the skewness of the
signed-intensity.

In summary, removing the directions with an anomalous signed-intensity
does not solve the observed discrepancy between the WMAP temperature
data and simulations. Consequently, taking the CMB temperature angular
power spectrum of the concordance cosmological model at face value,
we can conclude that the strong detection in the excess kurtosis of
the signed-intensity of the WMAP temperature data is not confined
to the directions with an anomalous signed-intensity.

\subsection{Discussion}

\label{sub:detection-stability}

Firstly, if the excess of kurtosis observed in the signed-intensity
of the WMAP temperature data is related to the CMB temperature field
itself (see Section \ref{sec:systematic}), the fact that this non-Gaussianity
is not confined to the directions with an anomalous signed-intensity
seems natural. Indeed, independently of the modification of the basic
inflationary scenario that might explain the non-Gaussianity of the
CMB temperature field \citep{bartolo04}, the cosmological principle
still implies its statistical isotropy, at least as a first approximation.
Non-Gaussian perturbations are therefore more naturally widely spread
over the whole sky. We also notice that the wavelet scales at which
the non-Gaussianity is observed are compatible with the size of CMB
anisotropies due to topological defects such as cosmic textures \citep{turok90},
or due to the Integrated Sachs-Wolfe effect, which is associated with
the time evolution of the gravitational potential of large scale structures\textbf{
}\citep{sachs67,rees68,martinez90}. Texture models suggest the presence
of a number of textures with angular sizes above $1^{\circ}$, which
can induce hot spots or cold spots of corresponding angular size in
the CMB \citep{turok90}. A recent Bayesian analysis \citep{cruz07b}
showed that the cold spot originally detected at $(\theta,\varphi)=(147^{\circ},209^{\circ})$
in galactic spherical coordinates with the axisymmetric Mexican hat
wavelet, is satisfactorily described by a texture with an angular
size on the celestial sphere around $10^{\circ}$. Other analyses
also showed that the time evolution of the gravitational potential
of large scale structures such as voids might induce cold spots in
the CMB with angular sizes of several degrees on the celestial sphere
\citep{martinez90}. The cold spot identified at $(\theta,\varphi)=(147^{\circ},209^{\circ})$
in galactic spherical coordinates with an angular size around $10^{\circ}$
could actually be explained in terms of a void at a redshift $z\simeq1$
and with a diameter around $300h^{-1}\,\textnormal{Mpc}$ \citep{inoue06,inoue07,rudnick}.

Secondly, as already emphasized, an excess of kurtosis in the wavelet
coefficient of the WMAP temperature data was previously detected with
the axisymmetric Mexican hat wavelet. The non-Gaussian deviation observed
with the axisymmetric Mexican hat wavelet is undoubtedly related to
the present detection in the excess kurtosis of the signed-intensity
with the 2GD wavelet, notably because it is observed with a similar
statistics at the same angular sizes on the celestial sphere. From
this point of view, both detections support one another.\textbf{ }But
the axisymmetric Mexican hat wavelet also allowed the detection of
the cold spot at $(\theta,\varphi)=(147^{\circ},209^{\circ})$ in
galactic spherical coordinates, which was interpreted to be the exclusive
origin of the excess of kurtosis detected \citep{cruz05}. On the
contrary, we have concluded that the detection in the excess kurtosis
of the signed-intensity of the WMAP temperature data with the 2GD
wavelet is not confined to the previously identified directions with
an anomalous signed-intensity. Consequently, even though the two detections
are similar, they probably simply do not identify the same non-Gaussian
content in the WMAP temperature data.

Finally, let us also underline that the values of the cosmological
parameters are affected by uncertainties associated with the limited
precision of measurement of the CMB temperature angular power spectrum.
These uncertainties are associated with the cosmic variance, but also
with systematic effects such as instrumental noise and residual foreground
emissions. The simulations produced for our analysis are obtained
from the angular power spectrum determined by the cosmological parameters
of the concordance model (\emph{i.e.} the three-year WMAP best-fit
model). They do not account for the quoted uncertainties. Consequently,
before giving full credit to our conclusions, a deep analysis should
be performed to check the stability of the various detections considered
when the WMAP temperature data are compared with simulations produced
from any possible angular power spectrum inside the experimental error
bars. Formally, any of our conclusions might be affected by this further
analysis, from the fact that the non-Gaussianity observed in the WMAP
temperature data is not confined to the directions with an anomalous
signed-intensity, up to the mere detection of an excess of kurtosis
in the signed-intensity. However, such an analysis would be very involved
and is not produced here. On the one hand, the stability of the detection
of an excess of kurtosis in the wavelet coefficient of the WMAP temperature
data with the axisymmetric Mexican hat wavelet was suggested \citep{vielva04}.
The same conclusion probably holds for the present detection of an
excess of kurtosis in the signed-intensity of the WMAP temperature
data with 2GD wavelet. On the other hand, the confinement analysis
itself is based on the previous detection of the distribution of directions
with an anomalous signed-intensity \citep{vielva07}. At present,
no analysis confirmed the stability of this distribution relative
to the uncertainties on the cosmological parameters. A possible excess
of power in the concordance model relative to the WMAP temperature
data \citep{spergel03,hinshaw07,monteserin07} might imply that a
part of the distribution on the celestial sphere of directions with
an anomalous signed-intensity was actually not detected. This would
probably not question the fact that this distribution is anomalous
at the wavelet scales between $a_{8}$ and $a_{11}$, corresponding
to angular sizes between $8.33^{\circ}$ and $13.3^{\circ}$, but
would simply suggest that the global significance level for the detection
was underestimated. However in such a case, the confinement analysis
itself, which explicitly requires the exclusion of all the extremal
values above a given threshold, both in the data and in the simulations,
might not be performed anymore. No conclusion relative to the possible
confinement of the non-Gaussianity observed to the directions with
an anomalous signed-intensity could therefore be reached.

\section{Conclusion}

\label{sec:conclusion}The decomposition of a signal on the sphere
with the steerable wavelet constructed from the second Gaussian derivative
gives access to morphological measures such as the orientation, signed-intensity,
and elongation of the signal's local features. In this work, the three-year
WMAP co-added data of the CMB temperature field are analyzed through
the first four statistical moments of the random fields associated
with these local morphological measures, at wavelet scales corresponding
to angular sizes between $27.5'$ and $30^{\circ}$ on the celestial
sphere. The statistical analysis is performed by comparison of the
data with ten thousand statistically isotropic and Gaussian simulations
produced from the concordance cosmological model. No detection is
made neither in the orientation analysis nor in the elongation analysis.
A strong detection is made in the excess kurtosis of the signed-intensity
of the WMAP data, with a significance level below $0.5\%$ at a wavelet
scale corresponding to an angular size around $10^{\circ}$, and confirmed
at neighbour scales. This supports a previous detection of an excess
of kurtosis in the wavelet coefficient of the WMAP data with the axisymmetric
Mexican hat wavelet. An isolated detection is also made in the variance
of the signed-intensity at the smallest wavelet scale. Systematic
effects such as residual point sources in the WMAP co-added CMB data
are suggested to originate this anomaly, which is consequently simply
discarded.

Instrumental noise and residual foreground emissions are not likely
to be at the origin of the detection in the excess kurtosis of the
signed-intensity. Large-scale modulations of the CMB related to some
unknown systematics are explicitly rejected as possible origins of
the detection. The observed non-Gaussianity is therefore probably
to be imputed to the CMB temperature field itself, thereby questioning
the basic inflationary scenario upon which the concordance cosmological
model relies. In this context, taking the CMB temperature angular
power spectrum of the concordance cosmological model at face value,
further analysis also naturally suggests that this non-Gaussianity
of the WMAP temperature data is not confined to the localized distribution
of anomalous signed-intensities. However, this last result, in particular,
might be sensitive to uncertainties affecting the cosmological parameters.
Further analyses should be performed before giving it full credit.

\section*{Acknowledgments}

The work of Y.W. is funded by the Swiss National Science Foundation
(SNF) under contract No. 200021-107478/1. Y.W. is also postdoctoral
researcher of the Belgian National Science Foundation (FNRS). The
work of P.V. is funded through an I3P contract from the Spanish National
Research Council (CSIC). P.V., R.B.B., and E.M.-G. are also supported
by the Spanish MCYT project ESP2004-07067-C03-01. The authors acknowledge
the use of the Legacy Archive for Microwave Background Data Analysis
(LAMBDA). Support for LAMBDA is provided by the NASA Office of Space
Science. The authors also acknowledge the use of the HEALPix and CAMB
softwares.

\label{lastpage}


\begin{thebibliography}{\protect\citeauthoryear{Mart\'inez-Gonz\'alez \& Sanz}{1990}}
\bibitem[\protect\citeauthoryear{Abramo et al.}{2006}]{abramo06}Abramo
L.R., Bernui A., Ferreira I.S., Villela T., Wuensche C.A., 2006, Phys.
Rev. D, 74, 063506

\bibitem[\protect\citeauthoryear{Bartolo et al.}{2004}]{bartolo04}Bartolo
N., Komatsu E., Matarrese S., Riotto A., 2004, Phys. Rep., 402, 103

\bibitem[\protect\citeauthoryear{Bennett et al.}{2003}]{bennett03}Bennett
C.L. et al., 2003, ApJS, 148, 1

\bibitem[\protect\citeauthoryear{Bernui et al.}{2006}]{bernui06}Bernui
A., Villela T., Wuensche C.A., Leonardi R., Ferreira I., 2006, A\&A,
454, 409

\bibitem[\protect\citeauthoryear{Bernui et al.}{2007}]{bernui07}Bernui
A., Mota B., Rebou\c{c}as M.J., Tavakol R., 2007, A\&A, 464, 479

\bibitem[\protect\citeauthoryear{Bielewicz et al.}{2005}]{bielewicz05}Bielewicz
P., Eriksen H.K., Banday A.J., Górski K.M., Lilje P.B., 2005, ApJ,
635, 750

\bibitem[\protect\citeauthoryear{Cay\'on et al.}{2005}]{cayon05}Cay\'on
L., Jin J., Treaster A., 2005, MNRAS, 362, 826

\bibitem[\protect\citeauthoryear{Chiang et al.}{2003}]{chiang03}Chiang
L.-Y., Naselsky P.D., Verkhodanov O.V., Way M.J., 2003, ApJ, 590,
L65

\bibitem[\protect\citeauthoryear{Chiang \& Naselsky}{2006}]{chiang06}Chiang
L.-Y., Naselsky P.D., 2006, Int. J. Mod. Phys D, 15, 1283

\bibitem[\protect\citeauthoryear{Coles et al.}{2004}]{coles04}Coles
P., Dinnen P., Earl J., Wright D., 2004, MNRAS, 350, 983 

\bibitem[\protect\citeauthoryear{Copi et al.}{2004}]{copi04}Copi
C.J., Huterer D., Starkman G.D., 2004, Phys. Rev. D, 70, 043515

\bibitem[\protect\citeauthoryear{Cruz et al.}{2005}]{cruz05}Cruz
M., Mart\'inez-Gonz\'alez E., Vielva P., Cayón L., 2005, MNRAS,
356, 29

\bibitem[\protect\citeauthoryear{Cruz et al.}{2006}]{cruz06}Cruz
M., Tucci M., Mart\'inez-Gonz\'alez E., Vielva P., 2006, MNRAS,
369, 57

\bibitem[\protect\citeauthoryear{Cruz et al.}{2007}]{cruz07a}Cruz
M., Cayón L., Mart\'inez-Gonz\'alez E., Vielva P., Jin J., 2007a,
ApJ, 655, 11

\bibitem[\protect\citeauthoryear{Cruz et al.}{2007}]{cruz07b}Cruz
M., Turok N., Vielva P., Mart\'inez-Gonz\'alez E., Hobson M.P.,
2007b, Science, 318, 1612

\bibitem[\protect\citeauthoryear{de Oliveira-Costa et al.}{2004}]{deOliveira04}de
Oliveira-Costa A., Tegmark M., Zaldarriaga M., Hamilton A., 2004,
Phys. Rev. D, 69, 063516

\bibitem[\protect\citeauthoryear{Donoghue \& Donoghue}{2005}]{donoghue05}Donoghue
E.P., Donoghue J.F., 2005, Phys. Rev. D, 71, 043002

\bibitem[\protect\citeauthoryear{Eriksen et al.}{2004a}]{eriksen04a}Eriksen
H.K., Hansen F.K., Banday A.J., Górski K.M., Lilje P.B., 2004a, ApJ,
605, 14

\bibitem[\protect\citeauthoryear{Eriksen et al.}{2004b}]{eriksen04b}Eriksen
H.K., Novikov D.I., Lilje P.B., Banday A.J., G\'orski K.M., 2004b,
ApJ, 612, 64

\bibitem[\protect\citeauthoryear{Eriksen et al.}{2005}]{eriksen05}Eriksen
H.K., Banday A.J., G\'orski K.M., Lilje P.B., 2005, ApJ, 622, 58

\bibitem[\protect\citeauthoryear{Eriksen et al.}{2007}]{eriksen07}Eriksen
H.K., Banday A.J., G\'orski K.M., Hansen F.K., Lilje P.B., 2007,
ApJ, 660, L81

\bibitem[\protect\citeauthoryear{Freeman et al.}{2006}]{freeman06}Freeman
P.E., Genovese C.R., Miller C.J., Nichol R.C., Wasserman L., 2006,
ApJ, 638, 1

\bibitem[\protect\citeauthoryear{Gordon}{2007}]{gordon07}Gordon C.,
2007, ApJ, 656, 636

\bibitem[\protect\citeauthoryear{G\'orski et al.}{2005}]{gorski05}G\'orski
K.M., Hivon E., Banday A.J., Wandelt B.D., Hansen F.K., Reinecke M.,
Bartelmann M., 2005, ApJ, 622, 759

\bibitem[\protect\citeauthoryear{Hansen et al.}{2004a}]{hansen04a}Hansen
F.K., Cabella P., Marinucci D., Vittorio N., 2004a, ApJ, 607, L67

\bibitem[\protect\citeauthoryear{Hansen et al.}{2004b}]{hansen04b}Hansen
F.K., Banday A.J., Górski K.M., 2004b, MNRAS, 354, 641

\bibitem[\protect\citeauthoryear{Helling et al.}{2006}]{helling06}Helling
R.C., Schupp P., Tesileanu T., 2006, Phys. Rev. D, 70, 063004

\bibitem[\protect\citeauthoryear{Hinshaw et al.}{2007}]{hinshaw07}Hinshaw
G. et al., 2007, ApJS, 170, 288

\bibitem[\protect\citeauthoryear{Inoue \& Silk}{2006}]{inoue06}Inoue
K.T., Silk J., 2006, ApJ, 648, 23

\bibitem[\protect\citeauthoryear{Inoue \& Silk}{2007}]{inoue07}Inoue
K.T., Silk J., 2007, ApJ, 664, 650

\bibitem[\protect\citeauthoryear{Katz \& Weeks}{2004}]{katz04}Katz
G., Weeks J., 2004, Phys. Rev. D, 70, 063527

\bibitem[\protect\citeauthoryear{Komatsu et al.}{2003}]{komatsu03}Komatsu
E. et al., 2003, ApJS, 148, 119

\bibitem[\protect\citeauthoryear{Land \& Magueijo}{2005a}]{land05a}Land
K., Magueijo J., 2005a, MNRAS, 357, 994

\bibitem[\protect\citeauthoryear{Land \& Magueijo}{2005b}]{land05b}Land
K., Magueijo J., 2005b, Phys. Rev. Lett., 95, 071301

\bibitem[\protect\citeauthoryear{Land \& Magueijo}{2007}]{land07}Land
K., Magueijo J., 2007, MNRAS, 378, 153

\bibitem[\protect\citeauthoryear{Larson \& Wandelt}{2004}]{larson04}Larson
D.L., Wandelt B.D., 2004, ApJ, 613, L85

\bibitem[\protect\citeauthoryear{Larson \& Wandelt}{2005}]{larson05}Larson
D.L., Wandelt B.D., 2005, preprint (arXiv:astro-ph/0505046v1)

\bibitem[\protect\citeauthoryear{L\'opez-Caniego et al.}{2007}]{lopezcaniego07}L\'opez-Caniego
M., Gonz\'alez-Nuevo J., Herranz D., Massardi M., Sanz J.L., De Zotti
G., Toffolati L, Argüeso F., 2007, ApJS, 170, 108

\bibitem[\protect\citeauthoryear{Mart\'inez-Gonz\'alez \& Sanz}{1990}]{martinez90}Mart\'inez-Gonz\'alez
E., Sanz J.L., 1990, MNRAS, 247, 473

\bibitem[\protect\citeauthoryear{McEwen et al.}{2005}]{mcewen05}McEwen
J.D., Hobson M.P., Lasenby A.N., Mortlock D.J., 2005, MNRAS, 259,
1583

\bibitem[\protect\citeauthoryear{McEwen et al.}{2006}]{mcewen06}McEwen
J.D., Hobson M.P., Lasenby A.N., Mortlock D.J., 2006, MNRAS, 371,
50

\bibitem[\protect\citeauthoryear{McEwen et al.}{2007}]{mcewen07}McEwen
J.D., Wiaux Y., Hobson M.P., Vandergheynst P., Lasenby A.N., 2007,
MNRAS, in press (arXiv:0704.0626v1 {[}astro-ph])

\bibitem[\protect\citeauthoryear{Monteser\'in et al.}{2007}]{monteserin07}Monteser\'in
C., Barreiro R.B., Vielva P., Mart\'inez-Gonz\'alez E., Hobson M.P.,
Lasenby A.N., 2007, preprint (arXiv:0706.4289v1 {[}astro-ph])

\bibitem[\protect\citeauthoryear{Mukherjee \& Wang}{2004}]{mukherjee04}Mukherjee
P., Wang Y., 2004, ApJ, 613, 51

\bibitem[\protect\citeauthoryear{Naselsky et al.}{2005}]{naselsky05}Naselsky
P.D., Chiang L.-Y., Olesen P., Novikov D.I., 2005, Phys. Rev. D, 72,
063512

\bibitem[\protect\citeauthoryear{Park}{2004}]{park04}Park C.-G.,
2004, MNRAS, 349, 313

\bibitem[\protect\citeauthoryear{Rees \& Sciama}{1968}]{rees68}Rees
M.J., Sciama D.W., 1968, Nature, 217, 511

\bibitem[\protect\citeauthoryear{Rudnick et al.}{2007}]{rudnick}Rudnick
L., Brown S., Williams L.R., 2007, 671, 40

\bibitem[\protect\citeauthoryear{Sachs \& Wolfe}{1967}]{sachs67}Sachs
R.K., Wolfe A.M., 1967, ApJ, 147, 73

\bibitem[\protect\citeauthoryear{Schwarz et al.}{2004}]{schwarz04}Schwarz
D.J., Starkman G.D., Huterer D., Copi C.J., 2004, Phys. Rev. Lett.,
93, 221301

\bibitem[\protect\citeauthoryear{Spergel et al.}{2003}]{spergel03}Spergel
D.N. et al., 2003, ApJS, 148, 175

\bibitem[\protect\citeauthoryear{Spergel et al.}{2007}]{spergel07}Spergel
D.N. et al., 2007, ApJS, 170, 377

\bibitem[\protect\citeauthoryear{Tojeiro et al.}{2006}]{tojeiro06}Tojeiro
R., Castro P.G., Heavens A.F., Gupta S., 2006, MNRAS, 365, 265

\bibitem[\protect\citeauthoryear{Turok \& Spergel}{1990}]{turok90}Turok
N., Spergel D.N., 1990, Phys. Rev. Lett. 64, 2736

\bibitem[\protect\citeauthoryear{Vielva et al.}{2004}]{vielva04}Vielva
P., Mart\'inez-Gonz\'alez E., Barreiro R.B., Sanz J.L., Cayón L.,
2004, ApJ, 609, 22

\bibitem[\protect\citeauthoryear{Vielva et al.}{2006}]{vielva06}Vielva
P., Wiaux Y., Mart\'inez-Gonz\'alez E., Vandergheynst P., 2006,
New Astron. Rev., 50, 880

\bibitem[\protect\citeauthoryear{Vielva et al.}{2007}]{vielva07}Vielva
P., Wiaux Y., Mart\'inez-Gonz\'alez E., Vandergheynst P., 2007,
MNRAS, 381, 932

\bibitem[\protect\citeauthoryear{Wiaux et al.}{2005}]{wiaux05}Wiaux
Y., Jacques L., Vandergheynst P., 2005, ApJ, 632, 5

\bibitem[\protect\citeauthoryear{Wiaux et al.}{2006a}]{wiaux06a}Wiaux
Y., Vielva P., Mart\'inez-Gonz\'alez E., Vandergheynst P., 2006a,
Phys. Rev. Lett., 96, 151303

\bibitem[\protect\citeauthoryear{Wiaux et al.}{2006b}]{wiaux06b}Wiaux
Y., Jacques L., Vielva P., Vandergheynst P., 2006b, ApJ, 652, 820
\end{thebibliography}
\end{document}